\documentclass[sigconf]{acmart}

\AtBeginDocument{%
  }

\usepackage[most]{tcolorbox}
\usepackage{xcolor}

\definecolor{themeRed}{HTML}{F25050}
\definecolor{themeBlue}{HTML}{506AF2}

\tcbset{
  highlights/.style={
    enhanced,
    colframe=blue!30,
    colback=blue!60,
    boxrule=0pt,
    corners=south,
    rounded corners=north,
    arc=1mm,
    boxsep=0pt,
    left=1mm,
    right=1mm,
    top=0.5mm,
    bottom=0.5mm, fontupper=\bfseries\color{white},
  },
}
\newtcolorbox{custombox}[1]{
	colback=gray!10,
	colframe=gray!70,
	left=1mm,
	right=1mm,
	top=1mm,
	bottom=1mm,
	fonttitle=\bfseries,
	arc=0mm,
	leftrule=1mm,
	rightrule=0mm,
	toprule=0mm,
	bottomrule=0mm,
	notitle,
	before=\par\smallskip\noindent,
	before upper={\textbf{#1: } },
}

\setcopyright{acmlicensed}
\copyrightyear{2024}
\acmYear{2024}
\setcopyright{rightsretained}
\acmConference[SIGCSE Virtual 2024]{Proceedings of the 2024 ACM Virtual
Global Computing Education Conference V. 1}{December 5--8, 2024}{Virtual
Event, NC, USA}
\acmBooktitle{Proceedings of the 2024 ACM Virtual Global Computing Education
Conference V. 1 (SIGCSE Virtual 2024), December 5--8, 2024, Virtual Event, NC,
USA}
\acmDOI{10.1145/3649165.3690100}
\acmISBN{979-8-4007-0598-4/24/12}

\begin{document}


\title[Synthetic Students: Bug Distributions Between LLMs and Computing Students]{Synthetic Students: A Comparative Study of Bug Distribution Between Large Language Models and Computing Students}


\author{Stephen MacNeil}
\affiliation{%
  \institution{Temple University}
  \city{Philadelphia}
  \state{PA}
  \country{US}}
\email{stephen.macneil@temple.edu	}
\orcid{0000-0003-2781-6619}

\author{Magdalena Rogalska}
\affiliation{%
  \institution{Temple University}
  \city{Philadelphia}
  \state{PA}
  \country{US}}
\email{m.rogalska@temple.edu}
\orcid{0009-0002-8876-3506}

\author{Juho Leinonen}
\affiliation{%
  \institution{Aalto University}
  \city{Espoo}
  \country{Finland}}
\email{juho.2.leinonen@aalto.fi}
\orcid{0000-0001-6829-9449}

\author{Paul Denny}
\affiliation{%
  \institution{University of Auckland}
  \city{Auckland}
  \country{New Zealand}}
\email{paul@cs.auckland.ac.nz}
\orcid{0000-0002-5150-9806}

\author{Arto Hellas}
\affiliation{
  \institution{Aalto University}
  \city{Espoo}
  \country{Finland}}
\email{arto.hellas@aalto.fi}
\orcid{0000-0001-6502-209X}

\author{Xandria Crosland}
\email{xcrosl1@wgu.edu}
\affiliation{%
 \institution{Western Governors University}
 \streetaddress{4001 S 700 E \#300}
 \city{Millcreek}
 \state{Utah}
 \country{USA}
 \postcode{84107}
}
\orcid{0009-0009-3720-8896}

\renewcommand{\shortauthors}{Stephen MacNeil et al.}

\begin{abstract}
Large language models (LLMs) present an exciting opportunity for generating synthetic classroom data.  Such data could include code containing a typical distribution of errors, simulated student behaviour to address the cold start problem when developing education tools, and synthetic user data when access to authentic data is restricted due to privacy reasons. In this research paper, we conduct a comparative study examining the distribution of bugs generated by LLMs in contrast to those produced by computing students. Leveraging data from two previous large-scale analyses of student-generated bugs, we investigate whether LLMs can be coaxed to exhibit bug patterns that are similar to authentic student bugs when prompted to inject errors into code. The results suggest that unguided, LLMs do not generate plausible error distributions, and many of the generated errors are unlikely to be generated by real students. However, with guidance including descriptions of common errors and typical frequencies, 
LLMs can be shepherded to generate realistic distributions of errors in synthetic code.
\end{abstract}






\maketitle

\section{Introduction}

Large language models (LLMs) present a promising new opportunity for generating synthetic data \cite{guo2024generative}, which may have important implications for computing education research and practice. Such data could include examples of code containing typical student errors, which could have useful practical applications.  For example, it provides a viable solution for conducting research in situations where access to authentic data is restricted due to privacy concerns~\cite{jones2020were}.  Teachers could also make use of such examples when developing learning resources, such as debugging tasks, or materials to help address common mistakes.  Having access to large amounts of synthetic data could also be useful for developers of educational tools.  For example, a well known problem when developing intelligent or adaptive tutoring systems is the `cold start problem' \cite{pankiewicz2021assessing}, where a lack of user data results in an initial mismatch between the system's internal model and a learner's actual performance.

Prior work outside of computing education contexts has begun to explore the use of LLMs for generating synthetic data.
%explored the goal of generating useful synthetic data, before and after the availability of LLMs.  
In the field of Human-Computer Interaction (HCI), Hämäläinen et al. generated synthetic questionnaire responses, demonstrating that LLMs can produce believable accounts of user experiences \cite{hamalainen2023evaluating}.  Although they show the potential for using synthetic data to ideate and pilot experiments, they suggest synthetic data should be validated with real data to ensure reliability. Similarly, Park et al. studied LLM-based agents to simulate human behavior and found interactions of the agents were human-like, and useful for designers~\cite{park2022social,park2023generative}.

Inspired by these developments in other research areas, in this work we investigate the potential of LLMs to generate synthetic code that mimics the distribution of bugs found in student-written code. The central research questions of this study are:

%reliably produce
\begin{itemize}
    \item \textbf{RQ1:} Can LLMs produce erroneous code upon request?
    \item \textbf{RQ2:} To what extent does directing an LLM through prompt engineering influence the distribution of bugs it generates?
    \item \textbf{RQ3:} How do bug distributions from an LLM correlate with or deviate from those generated by human students?
\end{itemize}

To address these questions, we compare the distribution of bugs generated by an LLM with those produced by computing students. Using publicly-available student data from previous studies, we investigate the effectiveness of different prompting strategies in guiding LLMs to generate realistic error distributions. Our findings suggest that while LLMs do not produce accurate distributions without guidance, they can do so with appropriate information about error frequencies.   Our work is the first to explore the use of LLMs for generating synthetic data for computing education research purposes, and we discuss avenues for future work.

\section{Related Work}
\subsection{Common Bugs in Students' Code} 

Learning to program involves developing an understanding of the syntax, structure and style of a programming language~\cite{luxton2018introductory}, and students encounter a wide range of syntax and logic errors during this process ~\cite{denny2011understanding, robins2006problem, altadmri201537}. Such errors include ``trivial mechanics'' errors such as syntax errors with braces, brackets, semicolons, and naming conventions~\cite{robins2006problem}, as well as errors related to the semantics of the learned language~\cite{altadmri201537}. Early research on errors in programming often centered on specific problems~\cite{soloway1982what, johnson1983bug, seppala2015we}. %soloway1983cognitive
Later, researchers increasingly used programming errors -- and programming process data -- in forming a deeper understanding of the problems ~\cite{ihantola2015educational}. As an early example of such work, Jadud~\cite{jadud2006methods} quantified the error fixing behavior of novice programmers, identifying a link between the error fixing behavior (or skill) and course outcomes.

In general, there are differences in the frequency of programming errors~\cite{spohrer1986novice, robins2006problem} and the time that it takes to fix such errors~\cite{denny2012all, brown2017novice}. % smith2019error, mccall2019new, jadud2006methods
The types of errors that students encounter also gradually change over time~\cite{altadmri201537, vihavainen2014novices}, and they can stem from multiple sources~\cite{altadmri201537,ettles2018common}. These sources include misinterpreting the programming problem and having flaws in programming knowledge~\cite{ettles2018common}, 
as well as 
%not to mention the potential 
the role of the programming language and the environment~\cite{kohn2019error, vihavainen2014novices, stefik2013empirical}. 

Research on programming errors has contributed to programming language design (e.g.~\cite{soloway1983cognitive}), and researchers have sought to help students with programming errors, for example, by improving programming error messages~\cite{becker2019compiler, denny2020error,leinonen2023using}. %becker2016effective, 
All such research builds on the availability of relevant data. However, although there are increasing number of programming datasets available~\cite{ihantola2015educational}, errors can vary between programming languages, and the distributions of errors may also differ between contexts.

\subsection{Generating Educational Content With LLMs} 

Large language models (LLMs), which are advanced transformer models, possess the ability to both comprehend and generate code and text. These capabilities enable LLMs to offer students personalized, just-in-time pedagogical support~\cite{denny2024desirable}
. For instance, LLMs have been utilized to enhance code comprehension by explaining code in plain English~\cite{macneil2023experiences, leinonen2023comparing,sarsa2022automatic} or by producing analogies to explain both code and underlying concepts like recursion~\cite{bernstein2024like, bernstein2024analyzing}. Leinonen et al. found that the explanations provided by LLMs were comparable in word length to those generated by peers in a classroom, though students rated the quality of LLM-generated explanations higher~\cite{leinonen2023comparing}. However, in other cases, LLMs have been shown to dramatically outperform students such as in identifying bugs~\cite{macneil2024decoding}. 
%{\color{red}This suggests that ...} 

In addition to supporting students with just-in-time content generation, some research has explored the creation of real-time content for instructors. For example, Tran et al. demonstrated that LLMs can generate multiple-choice questions with plausible distractors and correct answers based solely on the question stem~\cite{tran2023generating}. Doughty further extended this research by developing a pipeline for creating multiple-choice questions aligned with Bloom's Taxonomy~\cite{doughty2024comparative}. %{\color{red}Multiple authors have also demonstrated the capabilities for LLMs to generate code for students~\cite{kerslake2024stump}.} 
In their work, they observed that the pipeline resulted in similar learning objectives as those covered in the class.

What this prior work shows is that LLMs already appear to be capable of generating effective learning materials. In some cases, these materials are comparable to the ones that might be sourced from students~\cite{leinonen2023comparing} or crafted by instructors~\cite{tran2023generating, doughty2024comparative}. However, what remains unclear is how well these `similar' materials can be aligned to mimic the style and performance of students, including the common errors they might make.

\subsection{Synthetic Data Generation}

There are many reasons for generating synthetic data for research. Real data could be scarce~\cite{figueira2022survey}, hard to collect~\cite{hamalainen2023evaluating}, low quality~\cite{figueira2022survey}, or contain private information that cannot be shared (for example, medical records)~\cite{dankar2021fake}. One way of mitigating these issues is to try de-identify datasets~\cite{garfinkel2015identification}, although this can in some cases reduce the utility of the data~\cite{leinonen2017preventing}. These issues can be potentially sidestepped by generating synthetic data, if the quality of the generated data is similar to or greater than organic data.

Traditional approaches for generating synthetic data have included  algorithm-based approaches~\cite{sarsa2022empirical} %,piech2015deep,mannino2019real
and generative adversarial networks (GANs)~\cite{figueira2022survey}. %hernandez2022synthetic,
In education, synthetic data has been used, for example, to generate data to train and evaluate knowledge tracing methods~\cite{sarsa2022empirical} % piech2015deep
which aim to accurately model a learner's knowledge of the concepts they are practicing. 

%%% Paul: Note, I trimmed this section heavily, as these studies were already mentioned in the introduction.  The original is commented-out below
Recently, advancements in large language models (LLMs) have opened new possibilities for synthetic data generation. Research has explored various scenarios, including simulating social interactions. Park et al. discovered that interactions within a simulated community, termed `social simulacra,' can aid in prototyping, with simulated content often hard for experienced moderators to distinguish from real content ~\cite{park2022social}. In a follow-up study, they found that data generated by 25 LLM-based agents in a game-like environment was more believable than data from human crowdworkers~\cite{park2023generative}. Similarly, Hämäläinen et al. used GPT-3 to generate synthetic data on HCI experiences, particularly video games as art, finding the content human-like but less diverse than that created by humans~\cite{hamalainen2023evaluating}.

\section{Methods}

To evaluate our research questions, we conducted two comprehensive studies. In the first study, we aimed to replicate and extend the findings of Altadmri and Brown~\cite{altadmri201537}. Using their methodology, we employed GPT-4, a state-of-the-art LLM, to generate synthetic bugs. To address a lack of information about the precise programming problems used in their work, we conducted a second study where the programming problems were more explicitly defined. This second study replicates the bug frequencies identified by Rigby et al.~\cite{rigby2020miss} in their study of 900 programming students. An overview of the programs we used can be found in the virtual appendix\footnote{https://figshare.com/s/e0b0db319ca9ef73fa0c}.

\subsection{Prompting With Distributional Information} 
\label{sec:prompt-engineering}

\begin{figure}
    \centering 
    \includegraphics[width=\linewidth]{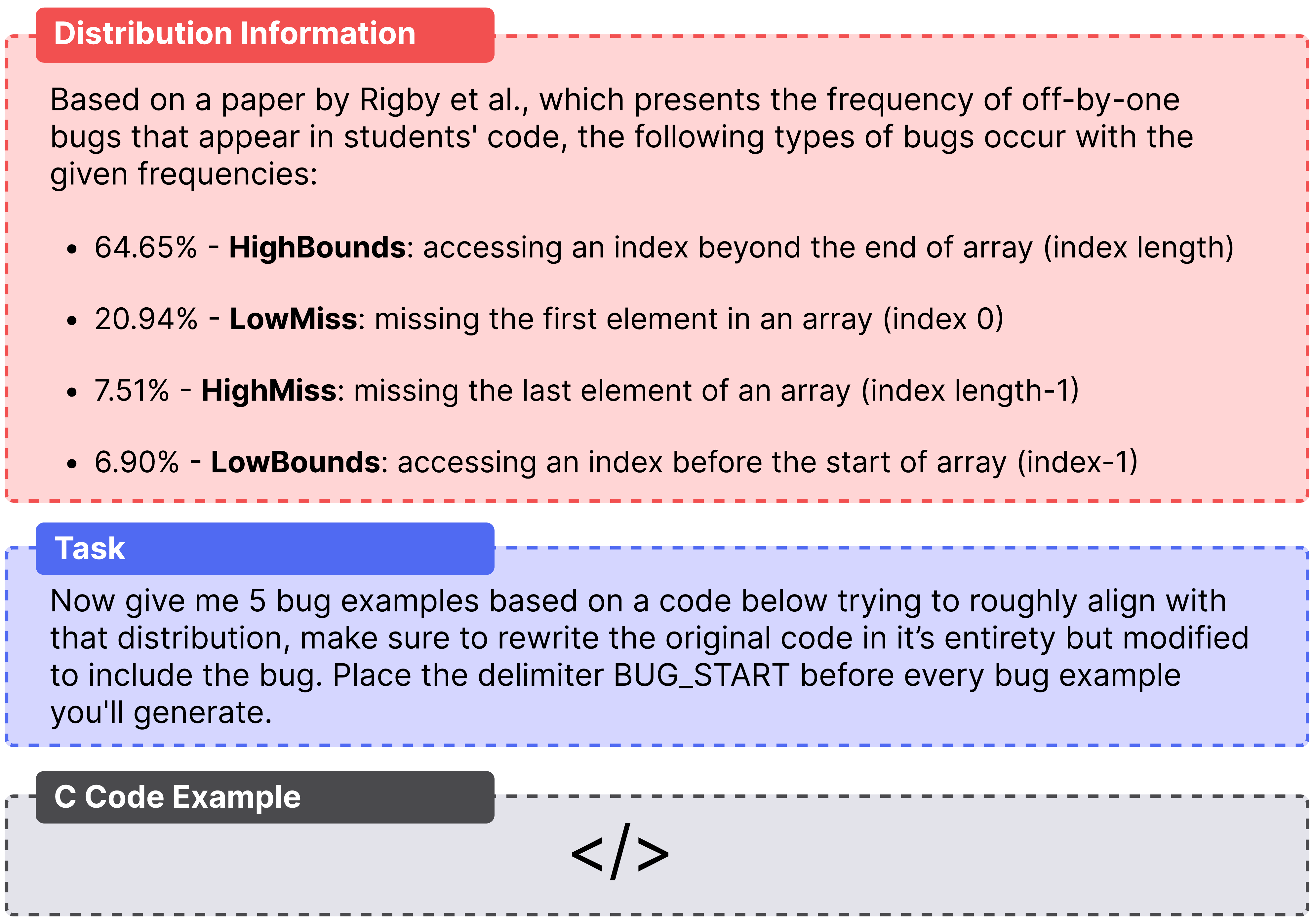}
    \caption{An overview of the prompts used in Study 2. The {\color{themeRed}\textit{Distributional Information}} was used in the \textit{Frequency-informed} and \textit{Taxonomy-informed} prompts. However, in the \textit{Taxonomy-informed} prompts, the specific frequency percentages were removed. The {\color{themeBlue}\textit{Task}} information and Code Example were used across all three prompts.}
    \label{fig:prompt-overview}
\end{figure}

To generate synthetic bugs, we investigated three prompting strategies that differed in the level of information provided about how students encounter similar problems in real-world contexts. An example based on Study 2 is presented in Figure~\ref{fig:prompt-overview}.

\subsubsection{Open-Ended Prompt} 

The baseline prompt was intentionally open-ended to gauge how closely LLMs align with student bug frequency. Here, no specific information about bug distribution was provided to the model.

\subsubsection{Taxonomy-Informed Prompt} 

In the second prompt, we explored the impact of providing basic information about the latent bug distribution. The model was given a list of bugs encountered by students in real-world contexts as reported by the two papers we replicated (i.e.: Altadmri and Brown~\cite{altadmri201537} and Rigby et al.~\cite{rigby2020miss}). This approach reflects how an instructor might have some intuition about common student bugs without precise frequency knowledge.

\subsubsection{Frequency-Informed Prompt}

In the final prompt, we provided the model with detailed information about the latent distribution of bug frequencies, including the specific frequencies at which students encountered each error. We used or computed the frequencies for each study based on the two papers being replicated~\cite{altadmri201537, rigby2020miss}.

\subsection{Study 1: Replication of Altadmri and Brown} 

In the first study, we conducted a replication of the work by Altadmri and Brown~\cite{altadmri201537}. Their research analyzed the frequency of bugs generated by real students across 37 million compilations. They identified 18 errors related to syntax, type, and semantics. These errors are summarized in Table~\ref{tab:study-1-bug-types}.% along with their shorthand abbreviations in Table~\ref{tab:study-1-bug-types}.

\begin{table}[!h]
%  \small
  \footnotesize
  \caption{Student Mistakes from Altadmri and Brown~\cite{altadmri201537}}
  \label{tab:study-1-bug-types}
  \begin{tabular}{cl}
    
    \toprule
    \textbf{Shorthand} & \textbf{Explanation of the Mistake} \\
    \midrule
    \multicolumn{2}{l}{\textit{Syntax errors:}} \\
    A & Confusing \texttt{=} with \texttt{==} \\ % Bug Code: A
    C & Mismatched parentheses \\ % Bug Code: C
    D & Confusing \texttt{\&} with \texttt{\&\&} \\ % Bug Code: D*
    E & Spurious semi-colon after \texttt{if}, \texttt{for}, \texttt{while}\\ % Bug Code: E*
    F & Wrong separator in \texttt{for} \\% Bug Code: F
    G & Wrong brackets in \texttt{if} \\ % Bug Code: G
    H & Using reserved keywords \\ % Bug Code: H
    J & Forgetting parentheses when calling methods \\ % Bug Code: J
    K & Spurious semi-colon after method header \\ % Bug Code: K
    L & Less-than / greater-than operators wrong \\ % Bug Code: L
    P & Including types in actual method arguments\\ % Bug Code: P
    \midrule
    \multicolumn{2}{l}{\textit{Type errors:}}\\ 
    I & Calling method with wrong types \\ % Bug Code: I
    Q & Type mismatch when assigning method result \\ % Bug Code: Q
    \midrule
    \multicolumn{2}{l}{\textit{Other semantic errors:}} \\
    B & Using \texttt{==} to compare strings \\ % Bug Code: B*
    M & Invoking instance method as static \\ % Bug Code: M
    N & Discarding method return \\ % Bug Code: N*
    O & Missing return statement \\ % Bug Code: O
    R & Missing methods when implementing interface \\ % Bug Code: R
    
  \bottomrule
\end{tabular}
\end{table}

To replicate their work, we chose five Java programs. Java was chosen because that was the language used by participants in their study~\cite{altadmri201537}. The five Java programs were chosen to be diverse; however, they do not perfectly match the range used in the prior study where the programming problems were not experimentally controlled and varied widely across the 37 million compilations.

We used GPT-4 to generate bugs and experimentally varied the prompts as described in Section~\ref{sec:prompt-engineering}. In total, 
%375 bugs were created by 
375 output programs with injected bugs
were created.  This was done by requesting the generation of five bugs for five code examples and repeating this process five times for each of the three prompt permutations to account for the probablistic nature of LLM responses, resulting in:

\texttt{3 prompts x 5 trials x 5 code examples x 5 bugs}

\subsection{Study 2: Replication of Rigby et al.}%~\cite{rigby2020miss}} 

The goal of the first study was to investigate whether LLMs are capable of generating similar distributions of bugs and syntax errors as students. However, it was challenging to replicate this prior work because they had investigated bug frequencies extracted from thousands of authentic programs which were solving a great variety of programming tasks. 

To more tightly control the programming tasks, we conducted a second study that replicates the work of Rigby et al.~\cite{rigby2020miss}. In their work, only four programming problems were used and the corresponding bug frequencies were computed based on more than 22,000 submissions from 900 students.  They focused purely on logic errors, in particular off-by-one errors for C code that iterates over an array.  They categorised the four mistakes that can cause an off-by-one error: missing the first element (index 0), missing the last element (index length-1), accessing an invalid index before the start (index -1), or accessing an index just past the end (index length).

Similar to Study 1, we used GPT-4 to generate bugs for the programming problems. We only modified the distributional information to align with the corresponding bugs and frequencies.

\subsection{Analyzing the LLM-Generated Bugs} 

\subsubsection{Deductive Coding and Inter-Rater Reliability} 

The first part of the analysis focused on deductively coding the generated data using the original taxonomies from each corresponding study~\cite{altadmri201537, rigby2020miss}. Two coders independently coded the data and we computed inter-rater reliability (IRR) using Cohen's Kappa to determine their agreement which is adjusted for class imbalances. For Study 1, these codes are listed in Table~\ref{tab:study-1-bug-types} and the Kappa for IRR between the two coders was 0.92. For Study 2, we used the same four bug types coded in their study, `High Bounds', `Low Bounds', `Low Miss', and `High Miss'. The Kappa for IRR between coders was 0.81.  

\subsubsection{Statistical Analyses}

The resulting frequency data was analyzed using the Chi Square goodness of fit test. Given that there were multiple distributions being compared, we corrected the critical p-values using the Bonferroni correction, which reduces Type I errors due to multiple comparisons. 

\subsubsection{Thematic Analysis of Out-of-Distribution Bugs}

When replicating  both studies~\cite{altadmri201537, rigby2020miss}, bug types that were not included in the original were coded `X'. These data were then analyzed using a thematic analysis approach to identify additional themes. The thematic analysis was guided by best practices~\cite{braun2006using} and followed a multi-step process with two coders analyzing the data independently but frequently discussing what they were observing and mediating their understanding. 

\section{Results}
\subsection{Study 1}

\begin{table}[]
%  \small
  \footnotesize
  \caption{Comparison of bug frequencies (\%) with Altadmri and Brown~\cite{altadmri201537}. The table includes out-of-distribution bugs from our thematic analysis and `-' denotes refusals.}
  \label{tab:study-results}
\begin{tabular}{clc|ccc}
\toprule
\textbf{Bug} & \textbf{Bug Type} & \textbf{Original} & \textbf{Frequency} & \textbf{Taxonomy} & \textbf{Baseline} \\
\midrule
C & Syntax & 33.1 & 17.6 & 13.6 & 4.8 \\
I & Type & 19.4 & 19.2 & 9.6 & 0 \\
O & Semantic & 14.3 & 17.6 & 12.8 & 4.8 \\
A & Syntax & 7.3 & 13.6 & 13.6 & 8.8 \\
N & Semantic & 5.1 & 5.6 & 2.4 & 0.8 \\
B & Semantic & 5.1 & 0.8 & 4.0 & 0.8 \\
M & Semantic & 3.6 & 4.0 & 2.4 & 1.6 \\
R & Semantic & 3.3 & 0 & 2.4 & 0 \\
P & Syntax & 2.2 & 0 & 4.0 & 0 \\
E & Syntax & 2.1 & 8.0 & 11.2 & 0.8 \\
K & Syntax & 1.6 & 0.8 & 3.2 & 0.8 \\
D & Syntax & 1.2 & 1.6 & 7.2 & 0 \\
J & Syntax & 0.8 & 0 & 0 & 0 \\
Q & Type & 0.7 & 0 & 0 & 1.6 \\
L & Syntax & 0.2 & 0 & 0.8 & 3.2 \\
F & Syntax & 0.1 & 0 & 0.8 & 0 \\
H & Syntax & <0.1 & 0 & 0 & 0 \\
G & Syntax & <0.1 & 0 & 0 & 0 \\
\midrule
    \multicolumn{3}{l}{\textit{Out-of-Distribution Errors}} \\
\midrule
- & None & N/A & 1.6 & 5.6 & 3.2 \\
X & Mixed & 0 & 4 & 0.8 & 12.8 \\
Y & Semantic & 0 & 0 & 0.8 & 19.2 \\
T & Type & 0 & 2.4 & 0 & 5.6 \\
W & Syntax & 0 & 0 & 0 & 7.2 \\
S & Semantic & 0 & 0.8 & 0 & 18.4 \\
U & Mixed & 0 & 2.4 & 4.8 & 5.6 \\

\bottomrule
\end{tabular}
\end{table}

\subsubsection{Generating Bugs With LLMs}

Our results suggest that LLMs can effectively create and integrate bugs into otherwise correct code. The refusal rates across prompts were extremely low (3.46\%) and this included instances where it returned correct code. Otherwise, the models were capable of producing and injecting bugs into the code. We did observe that the models often explicitly identified the bug in the code with a comment describing the bug.

\subsubsection{Bug Frequencies by Prompt Type}

Our results also show that providing the model with information about the distribution helped to ensure the distribution more closely matched the actual distribution of bugs generated by students. As shown in Table~\ref{tab:study-results}, the \textit{Baseline Prompt}, which had no information about the bug types or associated frequencies, produced code containing 68.8\% of bugs that were not in the original distribution. Conversely, the \textit{Frequency-informed} and \textit{Taxonomy-informed} prompts produced 9.6\% and 6.4\% of these out-of-distribution bugs. 

Based on a Chi Square Test, we observed that the \textit{Taxonomy-informed} ($\chi^2 = 8.7$, $p < 0.05$) and \textit{Baseline} ($\chi^2 = 17.7$, $p < 0.01$) prompts produced distributions that were statistically significantly different than the distributions present in the students' code. However, there was no significant difference for the \textit{Frequency-informed} prompt ($\chi^2 = 0.18$, $p = 0.91$). This suggests that the additional context was helpful in reproducing the original distribution.

Finally, across all three prompts, we observed a bias in the model where some bugs, such as A, E, and D were amplified by the model. For example, \textit{Bug E}, which is the error of adding a semi-colon after an if, for, or while statement, was common for both the \textit{Frequency-informed} (8.0\%) and \textit{Taxonomy-informed} (11.2\%) prompts despite being uncommon in the original student distributions (2.1\%)

%\tcbhighmath[highlights]{Bug E}

\begin{table}[]
    \centering
    \footnotesize
%    \small
    \caption{The themes of out-of-distribution bug types identified in our thematic analysis.}
    \label{tab:study1-out-of-distribution}
    \begin{tabular}{cp{7.5cm}}
        \toprule
       \textbf{Y} & \textit{Logic error:} Examples include counting 0 and 1 as primes, a function returning false in an if statement where it should return true, or forgetting to use a temporary variable while swapping variables. \\
        \textbf{T} & \textit{Type error:} A function or variable has the wrong type. For example, a function with return type \texttt{int} returns a string, or a variable of type \texttt{double} is assigned a char value. \\
        \textbf{W} & \textit{Undeclared or uninitialized variables:} Trying to use an undeclared variable or modify a variable that has not been assigned a value. \\
        \textbf{S} & \textit{Off-by-one error:} Starting a loop at 1 instead of 0, or iterating past the valid address in an array. \\  
        \textbf{U} & \textit{Operator confusion:} Confusing operators such as \texttt{\%, /, =, \&\&, ||}. For example, using \texttt{if (element \&\& toCheckValue)} instead of \texttt{if (element == toCheckValue)}. \\ 
        \bottomrule
    \end{tabular}
\end{table}

\subsubsection{Thematic Analysis of Out-of-Distribution Bugs}

In our initial coding, out-of-distribution bugs were labeled \textit{X}. We conducted a thematic analysis on these bugs to identify what types of bugs GPT-4 injected into the code. We identified five themes and additional bugs that did not fit into those themes. The five themes are described in Table~\ref{tab:study1-out-of-distribution}. The frequency of these five bugs are also reported in Table~\ref{tab:study-results}. Many of these errors were not compilation errors and were therefore not reported in the original study~\cite{altadmri201537}. 

\subsubsection{Examples of Out-of-Distribution Bugs}

In addition to the themes from the previous section, we also observed some `X' bugs that were less common but very interesting. For instance, when using the \textit{Baseline Prompt}, GPT-4 occasionally mixed syntax from different languages, such as confusing \texttt{.length()} with \texttt{.length}. Similarly, GPT-4 sometimes used \texttt{boolean} and \texttt{bool} interchangeably, and even misspelled it as \texttt{`bolean.'} These typos and syntactic confusions might reflect the types of errors students make when transitioning between programming languages \cite{denny2022novice}.

Some examples of out-of-distribution bugs were ones that would indicate considerable confusion if made by students (and thus might make useful teaching examples). 
%However, certain bugs seem unlikely to be made by students. 
For example, consider the following error produced from the \textit{Taxonomy-informed Prompt}: %where an assignment is mistakenly returned as an empty string:

\begin{quote}
    \texttt{{\color{themeRed}return} nstr = ""; {\color{gray}//Returning an assignment of empty string}}
\end{quote}

This bug is unusual because it conflates concepts such as return statements and variable assignments. Furthermore, assigning a variable and then immediately returning that variable is an unnecessary step as the empty string could just be returned directly. Moreover, in some languages this might return the memory address of the variable \texttt{nstr} (though in Java it would return the empty string) or produce a compilation error. Typically, programmers are taught to distinguish between assigning a value to a variable and returning a value from a function. 
%Returning an assignment, especially with an empty string, is not something many students would do. 

The \textit{Taxonomy-informed} and \textit{Frequency-informed} prompts seldom produced strange bugs. However, in one case the \textit{Frequency-informed} prompt included a string in the method signature: 

\begin{quote}
\texttt{public static String {\color{themeRed}reverse}({\color{themeRed}"Hello"}) \{}
\end{quote}

Finally, there was an instance in the \textit{Baseline} prompt where the model produced a conditional statement with the condition missing. 

\begin{quote}
    \texttt{{\color{themeRed}if} () \{ {\color{themeRed}return} {\color{themeBlue}true;} \}}
\end{quote}

Like the previous example, this error is unusual because it does not serve a functional purpose. Where other errors, such as using syntax from different languages interchangeably are mistakes students might make, it is more difficult to understand why a student would make such an error.
%unclear if students would write this code. 

\subsubsection{Refusal Rates}

While not common, we observed all three prompts resulted in the model refusing to add bugs to the code. The refusal rate was highest for the \textit{Taxonomy-informed} prompt (5.6\%) and lowest for the \textit{Frequency-informed} prompt (1.6\%). These differences are minor and likely driven largely by chance.  

%\subsubsection{Generating Rare Bugs} 

\subsection{Study 2}

The first study showed that providing information about the underlying distribution helped GPT-4 to replicate that distribution.  However, certain types of programming problems are more prone to specific bugs. For instance, off-by-one errors are highly unlikely in code that does not involve iteration. As a result, not being able to use the same programming problems as Altadmri and Brown~\cite{altadmri201537} was a limitation. To address this limitation, Study 2 focused on replicating prior work where only four programming problems (all consisting of iterating over an array) were attempted by every student in the study.  

\subsubsection{Bug Frequencies by Prompt Type}

Similar to Study 1, the \textit{Baseline} prompt produced more out-of-distribution errors (44\%) than the \textit{Frequency-informed} and \textit{Taxonomy-informed} prompts which only contained 6\% and 8\% of bugs that were not in the original distribution respectively. Unlike Study 1, the \textit{Baseline} prompt produced fewer out-of-distribution errors. This may be because the \textit{Baseline} prompt for Study 2 was constrained by only asking for `off-by-one' errors. 
We observed statistically significant differences between each distribution and the distribution of students' bugs. Based on Chi Square Tests (with p-values corrected using the Bonferroni correction), the \textit{Frequency-informed} ($\chi^2 = 69.9$, $p < 0.01$), \textit{Taxonomy-informed} ($\chi^2 = 115.7$, $p < 0.01$), and \textit{Baseline} ($\chi^2 = 76.4$, $p < 0.01$) prompts produced distributions of off-by-one errors that were different to those seen in practice from real students.  Manual inspection of the frequencies in Table \ref{tab:study-2-bug-types} indicate that the \textit{Frequency-informed} prompt produced a somewhat more realistic distribution, better matching the most common `HighBounds' error type.

\subsubsection{Examples of Out-of-Distribution Off-By-One Errors} 

We observed in the data many interesting errors that were injected by the LLM which were not strictly `off-by-one' errors (in the sense of loop iteration) but which did involve an adjustment (by 1) of a value or variable in the code.  
For instance, we observed bugs where the accumulator was decremented (or incremented) prior to it being returned by the function

\begin{lstlisting}
     return count-1;
\end{lstlisting}

We also observed some unusual errors, such as using a post-decrement operator within the loop condition, leading to quite subtle bugs (and which would cause the loop to terminate earlier than a typical off-by-one):
%.   Decrementing and incrementing in the same loop can lead to confusing behaviors:

\begin{lstlisting}
    for (int i = 0; i < n--; i++){
\end{lstlisting}

\subsubsection{Refusal Rates}

Refusals were uncommon (1\%--5\%) and similar to Study 1, with \textit{Frequency-informed} decreasing from Study 1 by 0.6\%, \textit{Taxonomy-informed} decreasing by 0.6\% and \textit{Baseline} decreasing by 0.2\%.

\begin{table}[]
  \footnotesize
%  \small
  \caption{Comparison of bug frequencies (\%) with Rigby et al.~\cite{rigby2020miss} where `-' and `X' represent refusals and out-of-distribution errors respectively.}
  \label{tab:study-2-bug-types}
\begin{tabular}{lc|ccc}
\toprule
\textbf{Bug} & \textbf{Original} & \textbf{Frequency} & \textbf{Taxonomy} & \textbf{Baseline}\\

\midrule
HighBounds & 64.6 & 30 & 21 & 20 \\
LowMiss & 20.9 & 21 & 21 & 13 \\
HighMiss & 7.5 & 19 & 21 & 0 \\
LowBounds & 6.9 & 21 & 20 & 19 \\
LM and HB & 0 & 2 & 2 & 1 \\ 
LM and HM & 0 & 0 & 2 &  0 \\
\midrule
    \multicolumn{3}{l}{\textit{Out-of-Distribution Errors}} \\
\midrule
- & N/A & 1 & 5 & 3 \\
X & 0 & 6 & 8 & 44 \\
\bottomrule
\end{tabular}
\end{table}

\section{Discussion}

In this paper, we explored whether LLMs can generate realistic synthetic bugs -- that is, that mirror the distribution of real bugs produced by students when working on programming problems.  Our results indicated that providing the model with some guidance helped considerably to align the generated bugs to those observed in practice.  In particular, providing a list of common bugs (\textit{Taxonomy-informed}) tended to improve the generated distribution over not including this information, whereas including frequency information as well (\textit{Frequency-informed}) provided even closer alignment.  In Study 1, we found that including frequency information helped the model to produce a corpus of buggy code with a distribution of errors that was statistically similar to the original data.

In both studies, we used the OpenAI API when making requests to the GPT-4 LLM, thus relying on the frequency information provided in our prompts across independent API calls.  In contrast, a chat-based LLM such as ChatGPT could take a more sophisticated approach by generating code (to select values at random according to a distribution) and executing it with its underlying code interpreter plug-in.  This would allow ChatGPT to produce a perfectly aligned distribution, limited only by its ability to generate correct bugs of each specified type, which is seems very capable of doing.  Of course, in practice, it may not be possible to accurately determine the probabilities (of bugs, or any other artefacts to be synthesized) in advance.  Even with data from prior research, as we had, such probabilities can be highly contextual and may vary from one programming task or educational setting to the next. For example, a previous replication of the original study by Brown and Altadmri with human students resulted in a slightly different distribution~\cite{ahadi2018learning}.  Nevertheless, our results demonstrate that LLMs have potential for generating synthetic data, such as bugs, using probability information when it is made available. 

Almost all of the 18 categories of errors in Study 1, and all four of the logic error categories in Study 2, had matching bug examples generated by the LLM.  This is quite impressive considering that the vast majority of code used in training LLMs is typically free from bugs, as code committed to public repositories is usually debugged beforehand.  Thus, the syntax and logic errors made by novices when learning to program are likely not well represented in the data available to LLMs when training.  However, especially for the baseline condition, some of the bugs produced were uncommon and in some cases unconventional. These bugs described as case studies in this paper were much more common for the baseline prompt which further highlights the value of including guidance to the models. Similarly, and consistent with prior work~\cite{armstrong2024silicone, navigli2023biases, kirk2021bias}, we observed biases in LLMs where some uncommon bugs were amplified by the model, even when given the frequency. 

\subsection{Toward Synthetic Students}

We see exciting avenues for future work exploring the use of LLMs to simulate individual students. For example, very recent work has shown that leveraging LLMs to simulate students answering MCQs can support item evaluation and help educators improve question quality \cite{lu2024generative}.  This suggests great potential for running simulations involving synthetic students.  A class of synthetic students, with different capability and error-proneness, could complete a proposed assessment to give feedback to the instructor on its suitability.  In theory, it may even be possible to to test interventions using synthetic students, allowing for experimentation in a controlled, risk-free environment before applying them in actual classroom settings. We elaborate on these possibilities in the following subsections: 

\subsubsection{Cold-Start Problem} 

The cold-start problem~\cite{schein2002methods} occurs when intelligent tutoring systems and autograders lack sufficient data to effectively assist students. This issue arises because these systems rely heavily on historical data to generate accurate recommendations and feedback. LLMs can help to supplant this need or to augment historical data if it is not sufficient. 

\subsubsection{Piloting In-Class Interventions}  

A persistent challenge in computing education research is the ethical concern of conducting interventions that might inadvertently harm students. One potential solution is to simulate student interactions within a classroom environment before implementing interventions. This approach allows researchers to test and refine their methods in a controlled, risk-free setting. While this strategy requires significant development and validation, it holds promise for improving the ethical standards of classroom research. Nonetheless, it is crucial that simulated approaches complement, rather than replace, in-class research involving actual students, as real-world student behavior cannot be fully captured by simulations. Simulations become another tool alongside pre-registration, informed consent, and power analysis.

\subsubsection{Predictors of Success} Building on prior computing education research focusing on predicting the success of students based on early performance~\cite{fincher2006predictors,hellas2018predicting}, one possible direction is to train LLM agents based on students in a class and then use those agents to identify students that are at risk of failing by simulating their performance through the rest of the course. 

Generating bugs similar to those encountered by students can also be beneficial for training TAs and instructors. This approach has been explored in the context of simulating students' responses to multiple-choice questions (MCQs)~\cite{lu2024generative} and training TAs by creating LLM agents that ask them questions about the assignments~\cite{markel2023gpteach}. Building on these efforts, LLMs could generate common bugs and coding design patterns, aiding TAs in identifying and addressing gaps in their knowledge before they begin working with students.

\section{Limitations} 

There are a few limitations to consider in this study. First, as mentioned in the discussion, there may be more deterministic methods for replicating the distribution of student data. The goal in this work was to understand the impact that information about the distribution has on model alignment. This is important because precise distributions are not always known and can vary based on the student population and course context~\cite{ahadi2018learning, altadmri201537, vihavainen2014novices}. This work also highlights that using LLMs to generate bugs without providing any information about an expected distribution will result predominately in irrelevant bugs. 

\section{Conclusions}

In this paper, we investigated the capabilities for LLMs to produce bugs with the same distribution as students in a classroom study. Across our two studies, we observed that giving the model information about the underlying distribution improved the ability of GPT-4 to produce relevant bugs. In cases where the distribution was not provided, GPT-4 was more likely to produce out-of-distribution bugs that in some cases would likely provide limited pedagogical benefit for students. Consequently, we propose the idea of synthetic students, which can mimic real student errors and behaviors, offering new opportunities for teacher training and practice.

\bibliographystyle{ACM-Reference-Format}
\bibliography{sample-base}

%\appendix

\end{document}